\documentclass[prc,preprint,showpacs,preprintnumbers,amsmath,amssymb]{revtex4}

\usepackage[mathscr]{eucal}
\usepackage{graphicx}
\def\slr#1{\setbox0=\hbox{$#1$}           
   \dimen0=\wd0                                 
   \setbox1=\hbox{/} \dimen1=\wd1               
   \ifdim\dimen0>\dimen1                        
      \rlap{\hbox to \dimen0{\hfil/\hfil}}      
      #1                                        
   \else                                        
      \rlap{\hbox to \dimen1{\hfil$#1$\hfil}}   
      /                                         
   \fi}

\def\be{\begin{eqnarray}}
\def\ee{\end{eqnarray}}

\begin{document}

\preprint{BCCNT: 04/071/326}

\title{Comparison of Models of Critical Opacity in the Quark-Gluon Plasma}

\author{Xiangdong Li}
\affiliation{%
Department of Computer System Technology\\
New York City College of Technology of the City University of New
York\\
Brooklyn, New York 11201 }%

\author{C. M. Shakin}
\email[email address:]{casbc@cunyvm.cuny.edu}

\affiliation{%
Department of Physics and Center for Nuclear Theory\\
Brooklyn College of the City University of New York\\
Brooklyn, New York 11210
}%

\date{July, 2004}

\begin{abstract}
In this work we discuss two methods of calculation of quark
propagation in the quark-gluon plasma. Both methods make use of
the Nambu-Jona-Lasinio model. The essential difference of these
calculations is the treatment of deconfinement. A model of
confinement is not included in the work of Gastineau, Blanquier
and Aichelin [hep-ph/0404207], however, the meson states they
consider are still bound for temperatures greater than the
deconfinement temperature $T_c$. On the other hand, our model
deals with unconfined quarks and includes a description of the
$q\overline{q}$ resonances found in lattice QCD studies that make
use of the maximum entropy method (MEM). We compare the
$q\overline{q}$ cross sections calculated in these models.

\end{abstract}

\pacs{12.39.Fe, 12.38.Aw, 14.65.Bt}

\maketitle


It is found in studies of heavy-ion interactions at RHIC that the
system very rapidly reaches equilibrium and that a model based
upon hydrodynamics is appropriate for the early stages of the
collision in which one expects to form a quark-gluon plasma [1].
Recently, Gastineau, Blanquier and Aichelin have proposed a model
of critical opacity based upon their calculation of the
$q\overline{q}$ interaction for temperatures in the range of $0$
to $350$ MeV [2]. A central feature of their model is the
temperature dependence of masses of the $u$, $d$ and $s$ quarks,
as well as the masses of the $\pi$ and $K$ mesons. (See Fig.\,1 of
Ref.\,[2].) In their model of the $q\overline{q}$ scattering
amplitude they calculate $s$, $t$ and $u$-channel exchange of
$\pi$ and $K$ mesons in the temperature range $0<T<350$ MeV. (We
remark that deconfinement takes place at $T_c\sim170$ MeV in
unquenched QCD calculations and at $T_c=270$ MeV in quenched
calculations.) In Ref.\,[2], the matrix elements in the $s$ and
$t$ channels are given by \be -i\mathcal{M}_t &=&
\delta_{c_1,c_3}\delta_{c_2,c_4}\overline{u}(p_3)Tu(p_1)[i\mathcal{D}^S_t(p_1-p_3)]v(p_4)T\overline{v}(p_2)
\\\nonumber &+&
\delta_{c_1,c_3}\delta_{c_2,c_4}\overline{u}(p_3)(i\gamma_5T)u(p_1)[i\mathcal{D}^P_t(p_1-p_3)]v(p_4)(i\gamma_5T)\overline{v}(p_2)
\ee

\be -i\mathcal{M}_s &=&
\delta_{c_1,c_2}\delta_{c_3,c_4}\overline{v}(p_2)Tu(p_1)[i\mathcal{D}^S_s(p_1+p_2)]v(p_4)T\overline{u}(p_3)
\\\nonumber &+&
\delta_{c_1,c_2}\delta_{c_3,c_4}\overline{v}(p_2)(i\gamma_5T)u(p_1)[i\mathcal{D}^P_s(p_1+p_2)]v(p_4)(i\gamma_5T)\overline{u}(p_3),
\ee where $p_1(p_2)$ is the momentum of the incoming
$q(\overline{q})$ and $p_3(p_4)$ that of the outgoing
$q(\overline{q})$. The $c_i$ are color indices and the various
$T$'s are the isospin projectors on the meson states. Here.
$\mathcal{D}^S$ and $\mathcal{D}^P$ are the meson propagators of
the form obtained in the NJL model, \be \mathcal{D}^S =
\frac{G_S}{1-G_SJ^S} \ee and \be \mathcal{D}^P =
\frac{G_P}{1-G_PJ^P}, \ee with $J^S$ and $J^P$ being the
polarization tensors in the scalar and pseudoscalar channels,
respectively. (In Eqs.\,(3) and (4) we have modified the notation
of Ref.\,[2] to be in closer correspondance to the notation we
have used in our work.)

In Ref.\,[2] the following Lagrangian is used without the term
proportional to $G_V$

\begin{flushleft}
\be \mathcal L&=&\overline{q}(i\slr\gamma-m^0)q+\frac{
G_S}{2}\sum_{i=0}^8 [(\overline{q} \lambda^{i} q)^2+(\overline{q}
i \gamma_5 \lambda^{i} q)^2]\\\nonumber &-&\frac{
G_V}{2}\sum_{i=0}^{8}[(\overline{q} \lambda^{i}\gamma_\mu
q)^2+(\overline{q} \lambda^{i}\gamma_5\gamma_\mu q)^2]\\\nonumber
&+&\frac{G_D}{2} \lbrace
\det[\overline{q}(1+\lambda_5)q]+\det[\overline{q}(1-\lambda_5)q]\rbrace
. \ee
\end{flushleft}

Here, $m^0$ is a current quark mass matrix, $m^0=diag(m_u^0,
m_d^0, m_s^0)$. The $\lambda_i$ are the Gell-Mann (flavor)
matrices and $\lambda^0=\sqrt{2/3}\mathbf{1}$, with $\mathbf{1}$
being the unit matrix. The fourth term is the 't Hooft
interaction. [Note that in the notation used in Ref.\,[2], $G_S$
and $G_V$ replace $G_S/2$ and $G_V/2$ of Eq.\,(5).]

In order to make contract with the results of lattice simulations
[3-6] we use the model with the number of flavors, $N_f$ =1.
Therefore, the $\lambda^i$ matrices in Eq.\,(5) may be replaced by
unity. We then have used

 \be \mathcal
L&=&\overline{q}(i\slr\gamma-m^0)q+\frac{G_S}{2}[(\overline{q}q)^2+(\overline{q}
i \gamma_5 q)^2]\\\nonumber &-&\frac{G_V}{2}[(\overline{q}
\gamma_\mu q)^2+(\overline{q}\gamma_5\gamma_\mu q)^2] \ee in order
to calculate the hadronic current correlation functions in earlier
work [7-9]. Spectral functions obtained using lattice QCD and MEM
are shown in Fig.\,1 and 2 [3].

\begin{figure}
\includegraphics[bb=0 0 280 235, angle=0, scale=1]{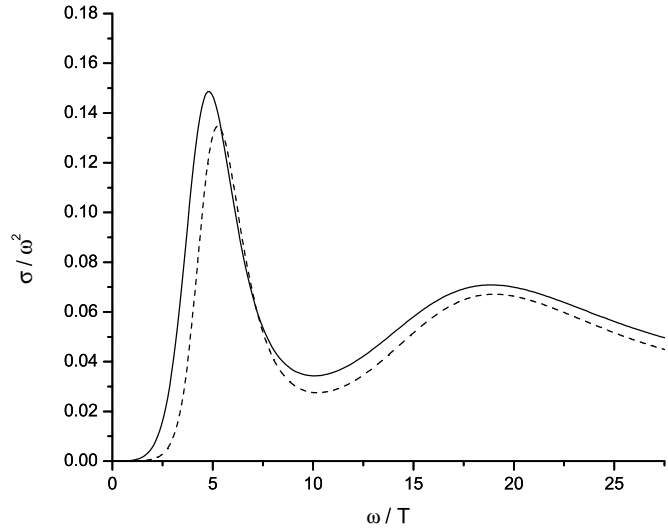}%
\caption{The spectral functions $\sigma/\omega^2$ for pseudoscalar
states obtained by MEM are shown [3]. The solid line is for
$T/T_c$ = 1.5 and the dashed line is for $T/T_c$ = 3.0. The second
peak is lattice artifact [3].}
\end{figure}

\begin{figure}
\includegraphics[bb=0 0 280 235, angle=0, scale=1]{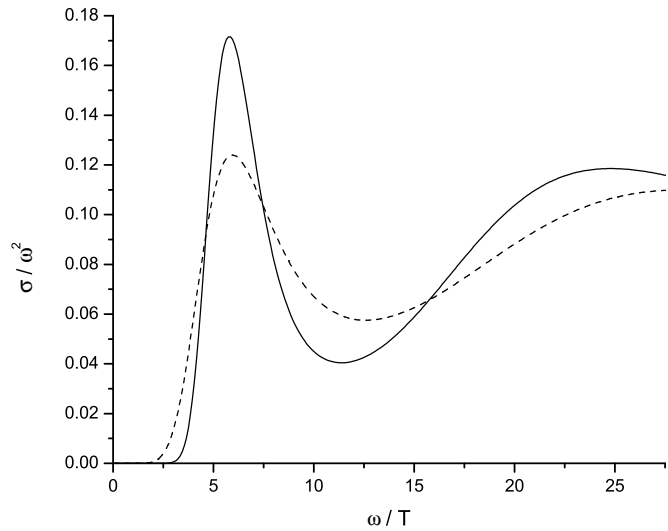}%
\caption{The spectral functions $\sigma/\omega^2$ for vector
states obtained by MEM are shown [3]. See the caption of Fig.\,1.
The second peak is lattice artifact [3].}
\end{figure}

\begin{figure}
\includegraphics[bb=0 0 280 235, angle=0, scale=1]{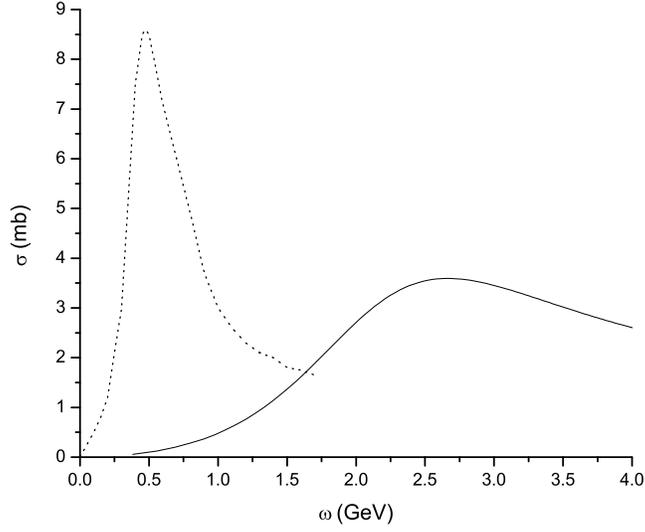}%
\caption{The dotted line shown the result for $T=350$ MeV obtained
in Ref.\,[2]. (See Fig.\,3 of Ref.\,[2] for this result and for
results at other temperatures.) The solid curve presents the
results calculated using our model at $T=350$ MeV [7-9].}
\end{figure}

\begin{figure}
\includegraphics[bb=0 0 280 235, angle=0, scale=1]{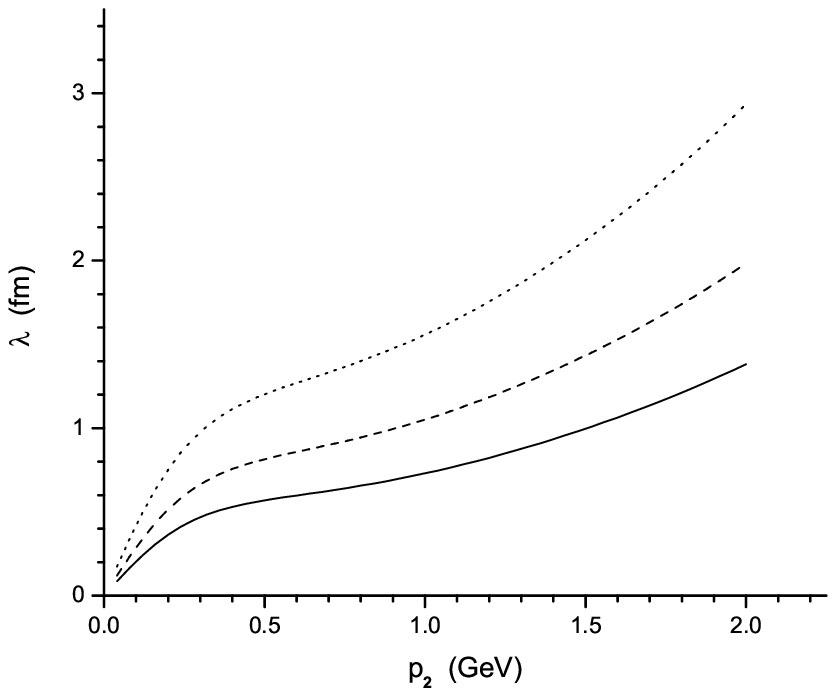}%
\caption{Quark mean-free path, $\lambda(p_2)$, shown for different
densities of antiquarks as a function of the quark momentum $p_2$.
(See Ref.\,[10] and Eq.\,(8) of the present work for the
parametrization of the antiquark densities used in our
calculation.) Here, $\mu=1.1$ GeV (dotted curve), $\mu=1.3$ GeV
(dashed curve) and $\mu=1.5$ GeV (solid curve).}
\end{figure}

We note that the unpolarized total cross section for a resonance
of mass $m_R$ in the $s$-channel may be written \be \sigma_{tot}
&=& \frac{1}{3\pi}E^2\frac{g^4}{(P^2-m_R^2)^2
+m^2_R\Gamma_R^2},\ee where $s=4E^2$ in the center-of-mass. The
more general result is obtained by using the expression \be
\frac{G_S}{1-G_SJ(P^2)} &\simeq&
-\frac{g^2}{P^2-m_R^2+im_R\Gamma_R}, \ee so that Eq.\,(7) becomes
\be \sigma_{tot} =
\frac{N}{3\pi}E^2\left|\frac{G_S}{1-G_SJ(P^2)}\right|^2 \ee where
$N$ is a statistical factor which we take to be $N=\sum_J(2J+1)
=8$, since we consider a sum over scalar, pseudoscalar, vector and
axial-vector resonances in our model. The calculation of $J(P^2)$
is discussed in great detail in our earlier work [7-9] and we do
not repeat that discussion here. (In that work we have used a
Gaussian regulator rather than the sharp cutoff usually used in
the case of the NJL model.)

The result for the cross section $u+\overline{u}\rightarrow
u+\overline{u}$ obtained in Ref.\,[2] is shown as a dotted line in
Fig.\,3. The solid line shows the result obtained in our model
[7-9].

The application of our model to calculate the quark mean-free path
is described in Ref.\,[10] and some results are shown in Fig.\,4
for different densities of antiquarks in the plasma [10]. That
distribution is parametrized as in Eq.\,(2.7) of Ref.\,[10] with
\be n(\overrightarrow{p_1}) &=& \frac{1}{\exp
\beta[E(\overrightarrow{p_1})-\mu]+1}, \ee $\beta=1/T$ and
$E(\overrightarrow{p_1})=[\overrightarrow{p_1}^2+m^2]^{1/2}$. (See
the caption of Fig.\,4.)

The main difference between our model and that of Ref.\,[2] is
that we recognize that the system is deconfined above the critical
temperature. The resonances calculated in Ref.\,[2] using the NJL
model that play the major role in the results of that work are not
present in the deconfined plasma. In our work we have used the
resonances found in the deconfined phase when using lattice QCD
and the maximum entropy method. Our work is closer in spirit to
that of the Stony Brook group [11-13] who also make reference to
the QCD lattice data in their discussion of the quite small quark
mean-free paths.



\end{document}